\documentclass[epj]{webofc}

\usepackage[varg]{txfonts}   

\begin{document}
\title{Warped extra dimension and inclined events at Pierre Auger Observatory}
\author{\firstname{Alexander} \lastname{Kisselev}\inst{1}\fnsep\thanks{\email{alexandre.kisselev@ihep.ru
}} \and \firstname{Liliya}
\lastname{Shkalina}\inst{2}\fnsep\thanks{\email{la.shkalina@physics.msu.ru
}} }
\institute{Institute for High Energy Physics, NRC ``Kurchatov
Institute'', 142281 Protvino, Russian Federation \and Department of
Physics, Moscow State University, 119991, Moscow, Russian
Federation}
\abstract{The generalized solution for the warp factor of the
Randall-Sundrum metric is presented which is symmetric with respect
to both branes and explicitly periodic in extra variable. Given that
the curvature of the 5-dimensional space-time is small, the expected
rate of neutrino-induced inclined events at the Surface Detector of
the Pierre Auger Observatory is calculated. Both the
``downward-going'' (DG) and ``Earth-skimming'' (ES) neutrinos are
considered. By comparing the expected event rate with the recent
Auger data on searching for neutrino candidates, the lower bound on
the fundamental gravity scale $M_5$ is obtained. The ratio of the
number of the ES air showers to the number of the DG showers is
estimated as a function of $M_5$.}
\maketitle

\section{Introduction}
\label{intro}

Ultra high energy (UHE) cosmic neutrinos play a key role in the
determination of the composition of the ultra high energy cosmic
rays (UHECRs) and their origin. UHE neutrinos are expected to be
produced in astrophysical sources in the decays of charged pions
created in the interactions of UHECRs with matter or radiation. They
can be also produced via interaction of the UHECRs with the cosmic
microwave background during propagation to the Earth (cosmogenic
neutrinos). UHE cosmic neutrinos are not deviated by magnetic fields
and could point back to their sources.

\begin{figure}[h]
\centering
\includegraphics[width=10cm,clip]{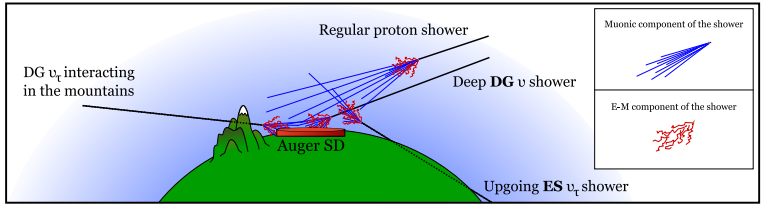}
\caption{Different types of showers induced by DG and ES neutrinos
(fig.~1 in \cite{Auger:2011}).}
\label{nu_events}
\end{figure}

Recently, three neutrinos of energy 1-2 PeV, as well as tens of
neutrinos above 10 TeV were detected with the IceCube experiment
\cite{IceCube:2014}. The cosmic neutrinos with energies near 1 EeV
are detectable with the Surface Detector (SD) of the Pierre Auger
Observatory (PAO) \cite{PAO}. In order to isolate neutrino-induced
events at the SD of the PAO, it is necessary to look for deeply
penetrating quasi-horizontal (inclined) air showers
\cite{Berezinsky:1969}-\cite{Zas:2005}. The PAO can efficiently
search for two types of neutrino-induced inclined air showers (see
fig.~\ref{nu_events}):
\begin{enumerate}
  \item
Downward-going (DG) neutrino-induced showers. They are initiated by
neutrinos moving with large zenith angle $\theta$ which interact in
the atmosphere close to the SD. At the PAO the search is restricted
to showers with $\theta > 60^\circ$ \cite{Auger:2015}. Note that the
background from hadronic showers above $10^{17}$ eV is
$\mathrm{O}(1)$ in 20 years, and it is negligible above $10^{19}$ eV
\cite{Anchordoqui:2010}.
  \item
Earth-skimming (ES) showers induced by upward tau neutrinos. They
can interact in the Earth's crust producing tau leptons. The latter
are efficiently produced at zenith angles $90^\circ < \theta <
95^\circ$ \cite{Auger:2015}. The tau leptons escape the Earth and
decay in the atmosphere close to the SD
\cite{Bertou:2002}-\cite{Feng:2002}.
\end{enumerate}

Recently, the Auger Collaboration reported on searches for DG
neutrinos in the zenith angle bins $60^\circ - 75^\circ$ and
$75^\circ - 90^\circ$, as well as for ES neutrinos. The data were
collected by the SD of the PAO from 1 January 2004 until 20 June
2013 \cite{Auger:2015}.%
\footnote{This search period is equivalent of 6.4 years of a
complete Auger SD working continuously \cite{Auger:2015}.}
No neutrino candidates were found. Assuming the diffuse flux of UHE
neutrinos to be $dN/dE_\nu = k E_\nu^{-2}$ in the energy range
$1.0\times10^{17}$ eV - $2.5\times10^{19}$ eV, the stringent limit
was obtained:
\begin{equation}\label{Auger_bound}
k < 6.4 \times 10^{-9} \mathrm{\ GeV \ cm^2 \ s^{-1} \ sr^{-1}} \;.
\end{equation}
This Auger limit is a factor 3.64 below the Waxman-Bachall bound on
neutrino production in optically thin astrophysical sources
\cite{WB:2001}:
\begin{equation}\label{WM_bound}
E_\nu^2 \frac{dN}{dE_\nu} = 2.33 \times 10^{-8} \mathrm{\ GeV \ cm^2
\ s^{-1} \ sr^{-1}} \;.
\end{equation}

In the Standard Model (SM) neutrino-nucleon cross sections are
expected to be very small in comparison with hadronic cross sections
even at UHEs \cite{Sarkar:2008}. That is why, the UHE cosmic
neutrinos can be regarded as unique probes of new interactions. In
the present paper a theory with an extra dimension (ED) is
considered to be a ``new physics'' theory. We will see that effects
coming from the ED can be significant or even dominant in the $\nu
N$-scattering at UHEs.

\section{General solution for the Randall-Sundrum metric}
\label{sec:RS}

In ref.~\cite{Randall:1999} the 5-dimensional space-time with
non-factorizable geometry was suggested as an alternative to the
models with flat extra dimensions. The Randall-Sundrum (RS) model
\cite{Randall:1999} predicts an existence of heavy Kaluza-Klein (KK)
excitations (massive gravitons). These massive KK resonances are
intensively searched for at the LHC.

The RS scenario is described by the following background warped
metric
\begin{equation}\label{RS_background_metric}
\quad ds^2 = e^{-2 \sigma (y)} \, \eta_{\mu \nu} \, dx^{\mu} \,
dx^{\nu} - dy^2 \;,
\end{equation}
where $\eta_{\mu\nu}$ is the Minkowski tensor with the signature
$(+,-,-,-)$, and $y$ is an extra coordinate. The periodicity
condition $y=y + 2\pi r_c$ is imposed, and the points $(x_\mu,y)$
and $(x_\mu,-y)$ are identified. Thus, one has a model of gravity in
the AdS$_5$ space-time compactified to the orbifold $S^1\!/Z_2$. The
orbifold has two fixed points, $y=0$ and $y=\pi r_c$. It is assumed
that there are two branes located at these points. All the SM fields
live on one of these branes.

The classical action of the RS scenario is given by
\cite{Randall:1999}
\begin{align}\label{action}
S &= \int \!\! d^4x \!\! \int_{-\pi r_c}^{\pi r_c} \!\! dy \,
\sqrt{G} \, (2 \bar{M}_5^3 \mathcal{R}
- \Lambda) \nonumber \\
&+ \int \!\! d^4x \sqrt{|g^{(1)}|} \, (\mathcal{L}_1 - \Lambda_1) +
\int \!\! d^4x \sqrt{|g^{(2)}|} \, (\mathcal{L}_2 - \Lambda_2) \;,
\end{align}
where $G_{MN}(x,y)$ is the 5-dimensional metric, with $M,N =
0,1,2,3,4$, $\mu = 0,1,2,3$. The quantities
\begin{equation}
g^{(1)}_{\mu\nu}(x) = G_{\mu\nu}(x, y=0) \;, \quad
g^{(2)}_{\mu\nu}(x) = G_{\mu\nu}(x, y=\pi r_c)
\end{equation}
are induced metrics on the branes, $\mathcal{L}_1$ and
$\mathcal{L}_2$ are brane Lagrangians, $G = \det(G_{MN})$, $g^{(i)}
= \det(g^{(i)}_{\mu\nu})$. $\bar{M}_5$ is the \emph{reduced}
5-dimensional Planck scale. The quantity $\Lambda$ is a
5-dimensional cosmological constant, while $\Lambda_{1,2}$ are brane
tensions.

\begin{figure}[h]
\centering
\includegraphics[width=8cm,clip]{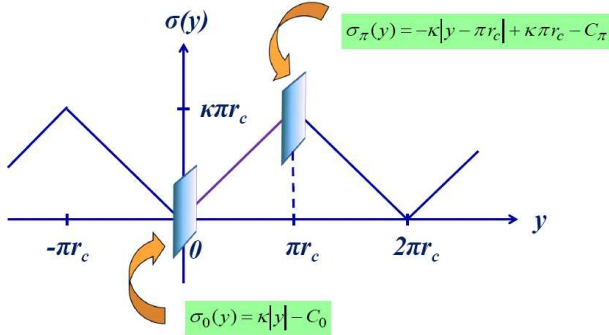}
\caption{Two equivalent solutions of Einstein-Hilbert's equations
for the function $\sigma(y)$ related to the different branes located
at $y=0$ and $y=\pi r_c$.}
\label{two_solutions}
\end{figure}

For the first time, the function $\sigma(y)$ was obtained in
\cite{Randall:1999} (see also the expression at the bottom of
fig.~\ref{two_solutions}):
\begin{equation}\label{sigma_1}
\sigma_0(y) = \kappa |y| - C_0 \;,
\end{equation}
where $\kappa$ is a parameter with a dimension of mass, which
defines the curvature of the 5-dimensional space-time.
This solution is consistent with the orbifold symmetry $y
\rightarrow - y$. However, it is attributed to the brane $y=0$ (see
fig.~\ref{two_solutions}), and it is
not symmetric with respect to the branes.%
\footnote{Note that in contrast to \eqref{sigma_1}, the RS1 solution
\cite{Randall:1999} has no constant term.}

Instead of \eqref{sigma_1}, one can use another \emph{equivalent}
solution related to the brane $y=\pi r_c$ (see the expression at the
top of fig.~\ref{two_solutions}):
\begin{equation}\label{sigma_2}
\sigma_\pi (y) =  -\kappa |y - \pi r_c| + |\kappa| \pi r_c - C_\pi
\;,
\end{equation}
In order to get an expression symmetric with respect to both branes,
we take half the sum of expressions \eqref{sigma_1} and
\eqref{sigma_2}. As a result, we come to the solution:
\begin{equation}\label{sigma_symmetric}
\sigma (y) = \frac{\kappa}{2} ( |y| - |y - \pi r_c| ) +
\frac{|\kappa| \pi r_c}{2} - C \;,
\end{equation}
where $C = (C_0 + C_\pi)/2$. The constant terms in
\eqref{sigma_symmetric} are chosen in such a way that one has
$\sigma (y) = \kappa y - C$ within the interval $0 < y < \pi r_c$.

Note that neither original RS1 solution nor generalized RS-like
solution \eqref{sigma_symmetric} obeys periodicity in $y$
\emph{explicitly}. One has to keep the periodicity condition in
``mind''. Fortunately, a new solution was recently proposed
\cite{Kisselev:2015} which is both symmetric with respect to the
branes and \emph{periodic} function of $y$:
\begin{equation}\label{sigma}
\sigma(y) = \frac{\kappa r_c }{2} \left[ \left| \mathrm{Arccos}
\left(\cos \frac{y}{r_c} \right) \right| - \left| \pi -
\mathrm{Arccos} \left(\cos \frac{y}{r_c} \right)\right| \right] +
\frac{\pi \,|\kappa| r_c }{2} - C \;,
\end{equation}
with the fine tuning relations:
\begin{equation}\label{fine_tuning}
\Lambda = -24 \bar{M}_5^3\kappa^2 \;, \quad \Lambda_1 = -
\,\Lambda_2 = 24 \bar{M}_5^3 \kappa \;.
\end{equation}
Here $\mathrm{Arccos(z)}$ is a \emph{principal value} of the
multivalued inverse trigonometric function $\arccos(z)$. It is
define by the inequality (see, for instance, \cite{Arccos}):
\begin{equation}\label{Arccos}
0 \leqslant \mathrm{Arccos}(z) \leqslant \pi \;, \quad -1 \leqslant
z \leqslant 1 \;.
\end{equation}
It means that \cite{Arccos}
\begin{equation}\label{arccos}
\mathrm{Arccos} (\cos x) = \left\{
\begin{array}{ll}
  x - 2n \pi , & 2n \pi \leqslant x \leqslant (2n +1)\pi \;, \\
  - x + 2(n +1)\pi  , & (2n +1)\pi \leqslant x \leqslant 2(n +1)\pi
  \;,
\end{array}
\right.
\end{equation}
where $n=0,1, \ldots$ It follows from \eqref{sigma}, (\ref{arccos})
that $\sigma(y) + C = \kappa y$ for $0 < y < \pi r_c$, $\sigma(y) +
C = \kappa (2\pi r_c - y)$ for $\pi r_c < y < 2\pi r_c$. and so on
(see fig.~\ref{two_solutions}).

Our solution \eqref{sigma} \\
- is symmetric with respect to the
branes: $\sigma(y)$ remains unchanged if
$y \rightarrow \pi r_c - y$, $k \rightarrow -k$;%
\footnote{Under the replacement $y \rightarrow \pi r_c - y$ the
positions of the branes are interchanged ($y=0 \rightleftarrows
y=\pi r_c$), while under the replacement $k \rightarrow -k$ their
tensions are interchanged ($\Lambda_1 \rightleftarrows \Lambda_2$).}
\\
- obeys the orbifold $Z_2$-symmetry: $\sigma(y) = \sigma(-y)$;
\\
- reproduces the jumps of $\sigma'(y)$ on both branes: $\sigma''(y)
= 2\kappa [\delta (y) - \delta (y - \pi r_c)]$, $0 \leqslant y
\leqslant \pi r_c$; \\
- is the periodic function of the extra coordinate: $\sigma(y + 2\pi
r_c n) = \sigma(y)$, $n=\pm 1, \pm 2, \ldots$ Note that the warped
function $\sigma(y)$ depends on the constant $C$.

\section{RS-like scenario with the small curvature}
\label{sec:RSSC}

By taking different values of $C$ in eq.~\eqref{sigma}, we come to
quite \emph{diverse physical scenarios} \cite{Kisselev:2015},
\cite{Kisselev:2016}. One of them ($C = 0$) is in fact the RS1 model
\cite{Randall:1999}. Another scheme ($C = \kappa \pi r_c$) describes
a geometry with a small curvature of five-dimensional space-time
(RSSC model) \cite{Giudice:2005}-\cite{Kisselev:2006}. In the RSSC
model the hierarchy relation looks like (we assume that $\kappa \pi
r_c \gg 1$)
\begin{equation}\label{hierarchy_relation_RSSC}
\bar{M}_{\mathrm{Pl}}^2  = \frac{\bar{M}_5^3}{\kappa} \, e^{2\kappa
\pi r_c} \;.
\end{equation}
So, one can take, for instance, $\kappa \sim 1$ GeV, $M_5 \sim 1$
TeV. Let us underline that the original RS scenario does not admit
the parameters to lie in this region, since $\kappa \sim \bar{M}_5
\sim M_{\mathrm{Pl}}$ TeV in it \cite{Randall:1999}.

The masses of the KK gravitons are proportional to the curvature
parameter $\kappa$:
\begin{equation}\label{m_n_2}
m_n = x_n \kappa \;,
\end{equation}
where $x_n$ are zeros of the Bessel function $J_1(x)$. The coupling
of the massive gravitons to the SM fields is given by the constant:
\begin{equation}\label{Lambda_pi_2}
\Lambda_\pi = \bar{M}_5 \left( \frac{\bar{M}_5}{\kappa}
\right)^{1/2} .
\end{equation}
Thus, the RSSC scenario predicts a spectrum of the KK gravitons
similar to that of the ADD model
\cite{Arkani-Hamed:1998}-\cite{Antoniadis:1998}. For the LHC
phenomenology of the RSSC model, see \cite{Kisselev:2008},
\cite{Kisselev:2013}.

\section{Neutrino-nucleon amplitude in the RSSC model}
\label{sec:nu_N}

For the UHE cosmic neutrinos, the scattering takes place in the
trans-Planckian kinematical region:
\begin{equation}\label{kinem_region}
\sqrt{s} \gtrsim \bar{M}_5 \gg -t \;,
\end{equation}
with $\sqrt{s}$ being the colliding energy and $t = - q_{\bot}^2$
four-dimensional momentum transfer. Remember that the inequality
$\kappa \ll \bar{M}_5$ is also satisfied. In the eikonal
approximation, which is valid in the kinematical region
\eqref{kinem_region}, the neutrino-proton amplitude is of the form:
\begin{equation}\label{ampl}
A_{\nu \rm p}(s, t) = 4 \pi i \, s \!\! \int\limits_0^{\infty} \! db
b J_0(b \, q_{\bot}) \left\{ 1 - \exp [i \chi(s, b)] \right\} \;.
\end{equation}
In its turn, the eikonal is given by the Fourier-Bessel
transformation of the eikonal:
\begin{equation}\label{eikonal}
\chi(s, b) = \frac{1}{4 \pi s} \! \int\limits_0^{\infty} \!
dq_{\bot} q_{\bot}  J_0(q_{\bot}  b) \, A_{\nu \rm p}^{\rm B} (s, t)
\;.
\end{equation}

The Born amplitude is given by the sum of gravi-Reggeons, i.e.
reggeized gravitons in the $t$-channel. Because of a presence of the
ED, the Regge trajectory of the graviton is splitting into an
infinite sequence of trajectories enumerated by the KK number
$n$~\cite{Kisselev:2004}:
\begin{equation}\label{trajectories}
\alpha_n(t) = 2 + \alpha_g' t  - \alpha_g' \, m_n^2, \quad n = 0, 1,
\ldots.
\end{equation}
In string theories, the slope of the gravi-Reggeons is universal,
and $\alpha_g' = M_s^{-2}$, where $M_s$ is the string scale. As a
result, in the RSSC model the gravity Born amplitude for the
neutrino scattering off \emph{a point-like} particle looks like
\cite{Kisselev:2005}, \cite{Kisselev:2009}:
\begin{equation}\label{Born_part_ampl}
A_{\rm grav}^{\mathrm{B}}(s, t) = \frac{\pi \alpha'_g s^2}{2
\Lambda_{\pi}^2} \sum_{n \neq 0} \left[ i - \cot\frac{\pi
\alpha_n(t)}{2} \right] \left( \frac{s}{\bar{M}_5}
\right)^{\alpha_n(t) -2} \;.
\end{equation}

The \emph{hadronic} Born amplitude in \eqref{eikonal} is defined by
a convolution of the gravity amplitude~\eqref{Born_part_ampl} and
skewed ($t$-dependent) PDFs $f_i(x,\mu^2; t)$:
\begin{equation}\label{Born_had_ampl}
A_{\nu \rm p}^{\rm B}(s, t) = \sum_{i=q,\bar{q},g} \int\limits_0^1
dx A_{\rm grav}^{\mathrm{B}}(xs, t) \, f_i(x,\mu^2; t) \;.
\end{equation}
We take the $t$-dependent PDFs in a factorized form:
\begin{equation}\label{skewed dist}
f_i(x,\mu^2; t) = f_i(x, \mu^2) D(t) \;.
\end{equation}
For the PDFs $f_i(x, \mu^2)$ we use the CT14 parametrization
\cite{CT14} and put $\mu^2 = |t|$. As for a suppression factor
$D(t)$, at hadron scales it was evaluated to be $D(t) =
\exp(tr_0^2)$, with $r_0^2 = 0.62 \mathrm{\ GeV}^{-2}$
\cite{Petrov:2002}. For larger $|t|$, we assume $D(t)$ to have a
power-like behavior. As a result, we come to the parametrization:
\begin{equation}\label{damping_factor}
D(t) =
\left\{
  \begin{array}{ll}
    \exp(t r_0^2) \;, & |t| \leqslant q_0^2 \;, \\
    \left(\mathrm{e} |t|/q_0^2 \right)^{-q_0^2 r_0^2} \;, & |t| > q_0^2 \;,
  \end{array}
\right.
\end{equation}
where $q_0 = m_\rho$ is the $\rho$-meson mass, and $\ln \mathrm{e} =
1$.

\begin{figure}[h]
\centering
\includegraphics[width=7cm,clip]{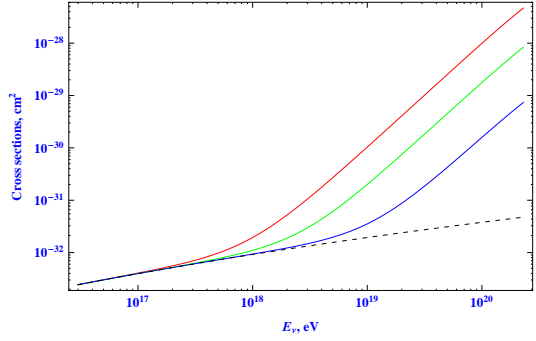}
\caption{Solid lines (from above): the neutrino total cross sections
for $\bar{M}_5 = 1.5$, 2.0 and 3.0 TeV. Dashed line: neutrino CC
total cross section.}
\label{cross_sections}
\end{figure}

The total cross sections with the account of the contribution from
the KK gravitons are shown in fig.~\ref{cross_sections}. We adopted
the SM $\nu N$ cross sections from \cite{Sarkar:2008}. One can see,
the  total cross section noticeably exceeds the SM one at energies
higher than $10^{18}$ or $10^{19}$ eV, depending on the reduced
5-dimensional Planck scale $\bar{M}_5$.

\section{Expected number of neutrino events at the PAO}
\label{sec:ev_num}

In \cite{Anchordoqui:2010} the following functional dependence of
the DG event rate on the new physics cross section
$\sigma_{\mathrm{NP}}$ was proposed for a bin in neutrino energy
($10^{18.5} \mathrm{\ eV} < E_\nu < 10^{19.5} \mathrm{\ eV}$):
\begin{equation}\label{DG:BSM_vs_SM_prel}
N_{\mathrm{BSM}}^{\mathrm{DG}} = N_{\mathrm{SM}}^{\mathrm{DG}} \,
\frac{\sigma_{\mathrm{CC}} +
\sigma_{\mathrm{NP}}}{\sigma_{\mathrm{CC}}} \;,
\end{equation}
where $N_{\mathrm{SM}}^{\mathrm{DG}}$ is the number of DG events in
the absence of new physics, $\sigma_{\mathrm{CC}}$ is a charged
current (CC) total cross section. $N_{\mathrm{BSM}}^{\mathrm{DG}}$
is the number of DG events with the account of interactions beyond
the SM. We propose a more general formula:
\begin{equation}\label{DG:BSM_vs_SM}
\mathcal{E}_{\mathrm{BSM}}^{\mathrm{DG}} (E_\nu) =
\mathcal{E}_{\mathrm{SM}}^{\mathrm{DG}} (E_\nu)\,
\frac{\sigma_{\mathrm{SM}}^{\mathrm{eff}}(E_\nu) +
\sigma_{\mathrm{NP}}(E_\nu)}{\sigma_{\mathrm{SM}}^{\mathrm{eff}}(E_\nu)}
\;,
\end{equation}
where $\mathcal{E}_{\mathrm{BSM}}^{\mathrm{DG}}$
($\mathcal{E}_{\mathrm{SM}}^{\mathrm{DG}}$) is the \emph{exposure}
of the SD of the PAO with (without) account of the new interaction.
In addition, instead of $\sigma_{\mathrm{CC}}$, an effective SM
cross section $\sigma_{\mathrm{SM}}^{\mathrm{eff}}$ is introduce in
\eqref{DG:BSM_vs_SM}:
\begin{equation}\label{sigma_eff}
\sigma_{\mathrm{SM}}^{\mathrm{eff}} = \sigma_{\mathrm{CC}}
\!\!\sum_{i=e,\mu,\tau} \! \, m_{\mathrm{CC}}^i +
3\sigma_{\mathrm{NC}} \, m_{\mathrm{NC}} + \sigma_{\mathrm{CC}}\,
m_{\mathrm{mount}} \;.
\end{equation}
Here $m_{\mathrm{CC}}^i$ and $m_{\mathrm{NC}}$ are relative mass
apertures for charged current and neutral current (NC) interactions
of the DG neutrinos at the PAO. The mass aperture
$m_{\mathrm{mount}}$ corresponds to the CC interaction of a $\tau$
neutrino within the mountains around the PAO (see
fig.~\ref{nu_events}). The relative mass apertures as functions of
the neutrino energy where calculated using data from Table~I of
ref.~\cite{Auger:2011}.%
\footnote{In particular, $\sum_{i=e,\mu,\tau} m_{\mathrm{CC}}^i$ is
equal to 0.69 (0.65) at $10^{18}$ ($10^{19}$) eV.}
Note that $\sum_{i=e,\mu,\tau} m_{\mathrm{CC}}^i + 3 m_{\mathrm{NC}}
+ m_{\mathrm{mount}} = 1$.

In contrast to the DG neutrino exposure, the exposure of the ES
neutrinos \emph{decreases} with the rise of the neutrino total
cross section:%
\footnote{Is is a generalization of the formula for the rate for ES
showers in the range $10^{18.5} \mathrm{\ eV} < E_\nu < 10^{19.5}
\mathrm{\ eV}$ \cite{Anchordoqui:2010}.}
\begin{equation}\label{ES:BSM_vs_SM}
\mathcal{E}_{\mathrm{BSM}}^{\mathrm{ES}}(E_\nu) =
\mathcal{E}_{\mathrm{SM}}^{\mathrm{ES}}(E_\nu) \,
\frac{\sigma_{\mathrm{CC}}^2(E_\nu)}{[\sigma_{\mathrm{CC}}(E_\nu) +
\sigma_{\mathrm{NP}}(E_\nu)]^2} \;.
\end{equation}

\begin{figure}[h]
\centering
\includegraphics[width=7cm,clip]{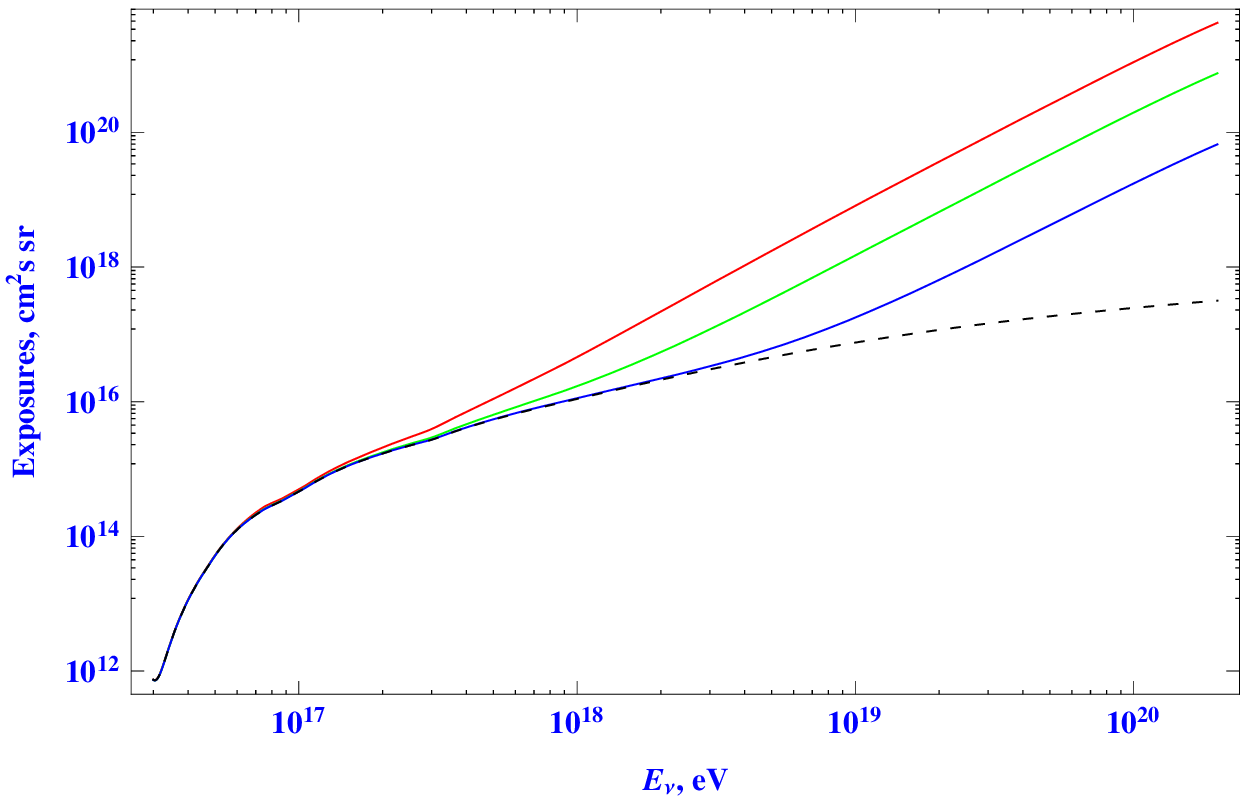}
\includegraphics[width=7cm,clip]{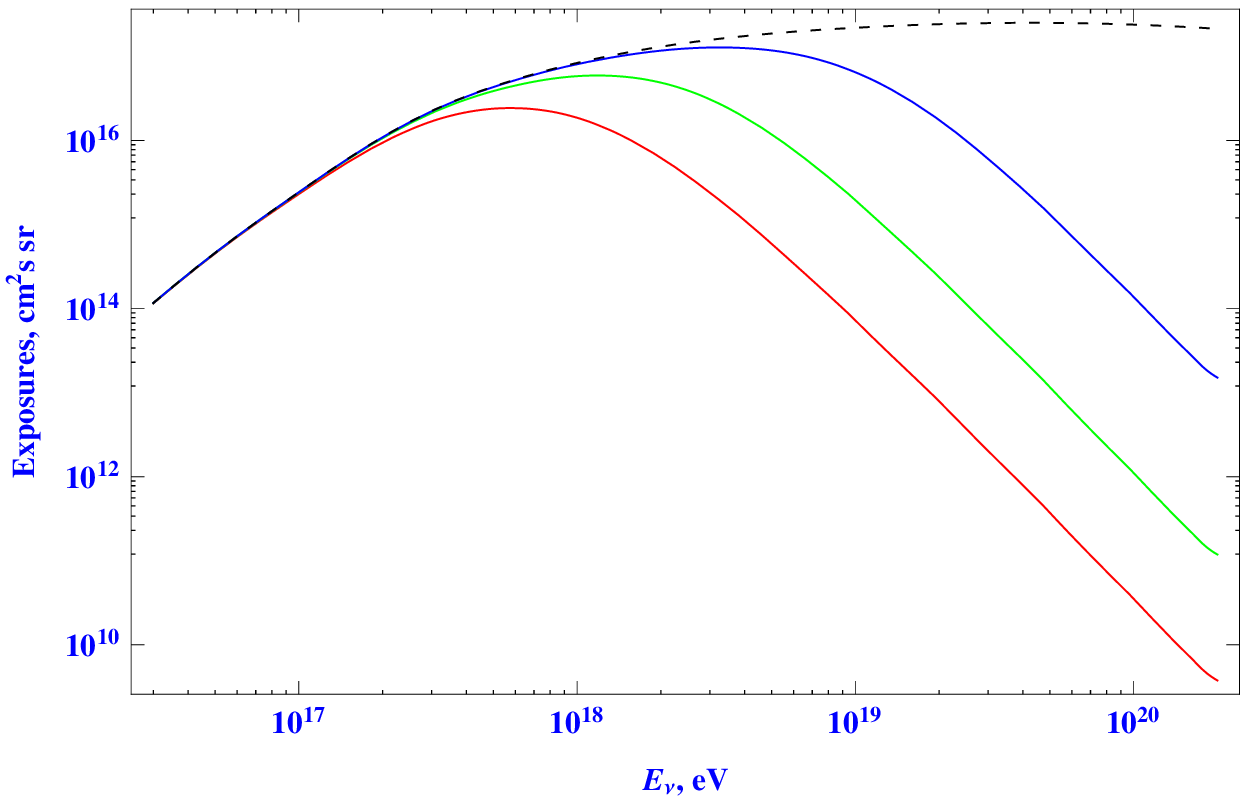}
\caption{Left panel: expected Auger exposures for the DG neutrinos
with zenith angle $75^\circ < \theta < 90^\circ$. Red, green and
blue line corresponds to $\bar{M}_5 = 1.5$, 2.0 and 3.0 TeV,
respectively. Dashed line: SM exposure taken from \cite{Auger:2015}.
Right panel: the same but for the ES neutrinos.}
\label{Auger_exposures}
\end{figure}

The formulas \eqref{DG:BSM_vs_SM} and \eqref{ES:BSM_vs_SM} allowed
us to calculate expected exposures of the SD of the PAO for the
period 1 January 2004 - 20 June 2013. The Auger data on exposures
for the SM neutrino interactions in the region from
$\log(E_\nu/\mathrm{eV}) = 17$ to 20.5 in steps of 0.5 were used
(see also fig.~3 from \cite{Auger:2015}). The results of our
calculations are presented in fig.~\ref{Auger_exposures}.

\begin{figure}[h]
\centering
\includegraphics[width=7cm,clip]{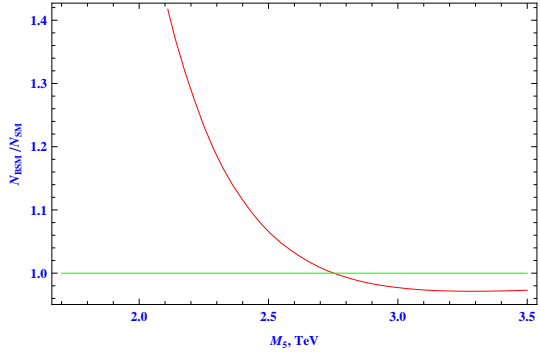}
\caption{The ratio of the neutrino events with and without effects
from EDs as a function of the gravity scale $\bar{M}_5$, expected at
the SD of the PAO. The event numbers correspond to the energy range
$1.0\times10^{17} - 2.5\times10^{19}$ eV.}
\label{BSM_vs_SM}
\end{figure}

The number of neutrino events is given by the formula:
\begin{equation}\label{numbers_events}
N_{\mathrm{ev}} = \int \frac{dN_\nu}{dE_\nu}(E_\nu) \,
\mathcal{E}(E_\nu) \, dE_\nu \;.
\end{equation}
We assume differential neutrino flux $dN/dE_\nu \sim E_\nu^{-2}$, as
well as a ratio $\nu_e : \nu_\mu : \nu_\tau = 1:1:1$. In
fig.~\ref{BSM_vs_SM} we compare the number of neutrino events
predicted in the RSSC model with the expected number of neutrino
events in the absence of a new physics. Both the DG events with
zenith angles $60^\circ < \theta < 90^\circ$ and ES events was taken
into account in the full energy region of the sensitivity of the SD
($1.0\times10^{17} - 2.5\times10^{19}$ eV).
As was already mentioned above, no neutrino events were seen at the
SD of the PAO. Thus, from fig.~\ref{BSM_vs_SM} we obtain the lower
bound on 5-dimensional \emph{reduced} Planck scale:
\begin{equation}\label{M5_bound}
\bar{M}_5 > 2.75 \mathrm{\ TeV} \;.
\end{equation}
Correspondingly, for the 5-dimensional  Planck scale we find $M_5 >
2.75 \times (2\pi)^{1/3} = 5.07$ TeV.

\begin{figure}[h]
\centering
\includegraphics[width=7cm,clip]{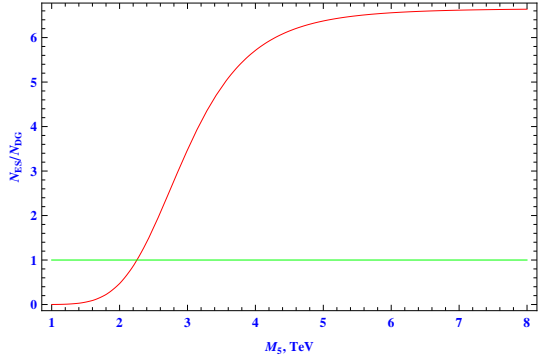}
\caption{The expected ratio of the ES neutrinos to the DG neutrinos
(with zenith angle $75^\circ < \theta < 90^\circ$) at the SD of the
PAO as a function of the gravity scale $M_5$.}
\label{ES_vs_DG}
\end{figure}

We have also calculated an expected ratio of the ES neutrino showers
to the DG neutrino showers with the zenith angles $75^\circ < \theta
< 90^\circ$ (see fig.~\ref{ES_vs_DG}).
We predict $N_{\mathrm{ES}}/N_{\mathrm{DG}}$ to be 2.8 (5.7) for
$\bar{M}_5 = 2.8$ (4.0) TeV. For the SM interactions, this ratio was
estimated by the Auger Collaboration to be equal to 6.0
\cite{Auger:2015}.

\section{Conclusions}

In the present paper we have studied the neutrino-induced inclined
(quasi-horizontal) events at the Surface Detector of the Pierre
Auger Observatory in the Randall-Sundrum scenario with the extra
dimension and warped metric. We have presented the general solution
for the metric \eqref{sigma} which is symmetric with respect to both
branes and explicitly periodic in extra variable. In the framework
of the RS-like model with the small curvature of the 5-dimensional
space-time, the exposures of the SD of the PAO for the downward
going and Earth-skimming neutrinos are estimated
(fig.~\ref{Auger_exposures}). The lower bound on the fundamental
gravity scale $\bar{M}_5$ is obtained \eqref{M5_bound}. The ratio of
the number of the ES air showers to the number of the DG showers,
$N_{\mathrm{ES}}/N_{\mathrm{DG}}$, is calculated as a function of
$\bar{M}_5$ (fig.~\ref{ES_vs_DG}).

\section*{Acknowledgements}

The authors are indebted to J. Alvarez-Mu\~{n}iz for sending
numerical data on the Auger exposures shown in fig.~3 from
ref.~\cite{Auger:2015}.


\begin{thebibliography}{99}
%
\bibitem{IceCube:2014}
M.G.~Aartsen \emph{et al.} (IceCube Collab.), Phys. Rev. Lett.
\textbf{113}, 101101 (2014)
%
\bibitem{PAO}
J.~Abraham \emph{et al.} (Auger Collab.), Nucl. Instrum. Meth. A
\textbf{523}, 50 (2004)
%
\bibitem{Berezinsky:1969}
V.S.~Berezinsky and G.T.~Zatsepin, Phys. Lett. B \textbf{28}, 423
(1969); V.S.~Berezinsky and A.Yu.~Smirnov, Astrophys. Space Sience
\textbf{32}, 461 (1975)
%
\bibitem{Capelle:1998}
K.S.~Capelle, J.W.~Cronin, G.~Parente and E.~Zas, Astropart. Phys.
\textbf{8}, 321 (1998)
%
\bibitem{Zas:2005}
E.~Zas, New J. Phys., \textbf{7}, 130 (2005)
%
\bibitem{Auger:2015}
A.~Aab \emph{et al.} (Auger Collab.), Phys. Rev. D \textbf{91},
092008 (2015)
%
\bibitem{Anchordoqui:2010}
L.~Anchordoqui, H.~Goldberg, D.~G\'{o}ra \emph{et al.}, Phys. Rev. D
\textbf{82}, 043001 (2010)
%
\bibitem{Bertou:2002}
X.~Bertou et al., Astropart. Phys. \textbf{17}, 183 (2002)
%
\bibitem{Feng:2002}
J.L.~Feng, P.~Fisher, F.~Wilczek and T.M.~Yu, Phys. Rev. Lett.
\textbf{88}, 161102 (2002)
%
\bibitem{Auger:2011}
P.~Abreu \emph{et al.} (Auger Collab.), Phys. Rev. D \textbf{84},
122005 (2011)
%
\bibitem{WB:2001}
E.~Waxman and J.N.~Bachall, Phys. Rev. D \textbf{64}, 023002 (2001)
%
\bibitem{Sarkar:2008}
A.~Cooper-Sarkar and S.~Sarkar, JHEP \textbf{0801}, 075 (2008)
%
\bibitem{Randall:1999}
L.~Randall and R.~Sundrum, Phys. Rev. Lett. \textbf{83}, 3370 (1999)
%
\bibitem{Kisselev:2015}
A.V.~Kisselev, Proc. of the Third Annual Large Hadron Collider
Physics Conference (LHCP2015), August 31-September 2, 2016, St.
Petersburg, Russia
%
\bibitem{Arccos}
I.S.~Gradshteyn and I.M.~Ryzhik, \emph{Table of Integrals, Series,
and Products, Eighth Edition} (Eds. D.~Zwillinger and V.~Moll,
Academic Press, US, 2014) 55-56.
%
\bibitem{Kisselev:2016}
A.V.~Kisselev, \emph{Generalization of the Randall-Sundrum
solution}, arXiv:1512.01091, Nucl. Phys. B \textbf{909} (2016), to
appear.

%
\bibitem{Giudice:2005}
G.~F.~Giudice, T.~Plehn and A.~Strumia, Nucl. Phys. B \textbf{706},
455 (2005)
%
\bibitem{Kisselev:2005}
A.V.~Kisselev and V.A.~Petrov, Phys. Rev. D \textbf{71}, 124032
(2005)
%
\bibitem{Kisselev:2006}
A.V.~Kisselev, Phys. Rev. \textbf{D 73}, 024007 (2006)
%
\bibitem{Arkani-Hamed:1998}
N.~Arkani-Hamed, S.~Dimopoulos and G.~Dvali, Phys. Lett. B
\textbf{429}, 263 (1998); Phys. Rev. D \textbf{59}, 086004 (1999)
%
\bibitem{Antoniadis:1998}
I.~Antoniadis, N.~Arkani-Hamed, S.~Dimopoulos and G.~Dvali, Phys.
Lett. B \textbf{436}, 257 (1998)
%
\bibitem{Kisselev:2008}
A.V.~Kisselev, JHEP \textbf{0809}, 039 (2008)
%
\bibitem{Kisselev:2013}
A.V.~Kisselev, JHEP \textbf{1304}, 025 (2013)
%
\bibitem{Kisselev:2004}
A.V.~Kisselev and V.A.~Petrov, Eur. Phys. J. C \textbf{36}, 103
(2004); \emph{ibid} \textbf{37}, 241 (2004)
%
\bibitem{Kisselev:2009}
A.V.~Kisselev, Open Astron. J. \textbf{2}, 12 (2009)
%
\bibitem{CT14}
S.~Dulat \emph{et al.},  Phys. Rev. D \textbf{93}, 033006 (2016)
%
\bibitem{Petrov:2002}
V.A.~Petrov and A.V.~Prokudin, Eur. Phys. J. C, \textbf{23}, 135
(2002)
%
\end{thebibliography}
\end{document}